\documentclass[aps, prb, reprint, superscriptaddress, showpacs]{revtex4-1}
\usepackage{color, graphicx}
\usepackage{amssymb,amsmath, amsfonts, bm}
\usepackage[colorlinks,citecolor=blue, urlcolor=blue, linkcolor=blue]{hyperref}

\begin{document}
\title{Efficient treatment of the high-frequency tail of the
  self-energy function and its relevance for multiorbital models} 

\author{Gang Li}
\email[]{gangli@physik.uni-wuerzburg.de}
\author{Werner Hanke}
\affiliation{Institut f\"ur Theoretische Physik und Astrophysik,
  Universit\"at W\"urzburg, 97074 W\"urzburg, Germany} 


\begin{abstract}
  In this paper, we present an efficient and stable method to
  determine the one-particle Green's function in the
  hybridization-expansion continuous-time (CT-HYB) quantum Monte Carlo
  method, within the framework of the dynamical mean-field theory
  (DMFT).  
  The high-frequency tail of the impurity self-energy is replaced with a
  noise-free function determined by a dual-expansion around the atomic
  limit.  
  This method does not depend on the explicit form of the interaction term. 
  More advantageous, it does not introduce any additional numerical
  cost to the CT-HYB simulation.  
  We discuss the symmetries of the two-particle vertex, which can be
  used to optimize the simulation of the four-point correlation functions
  in the CT-HYB. 
  Here, we adopt it to accelerate the dual-expansion calculation, which
  turns out to be especially suitable for the study of material
  systems with complicated band structures. 
  As an application, a two-orbital Anderson impurity model with a general
  on-site interaction form is studied.
  The phase diagram is extracted as a function of the Coulomb
  interactions for two different Hund's coupling strengths.
  In the presence of the hybridization between different orbitals, for
  smaller interaction strengths,        
  this model shows a transition from metal to band-insulator.  
  Increasing the interaction strengths, this transition is replaced by
  a crossover from Mott-insulator to band-insulator behavior. 
\end{abstract}

\maketitle

\section{Introduction}
The study of electronic structure of transition metal and
heavy-fermion materials is one of the most active fields in
condensed-matter physics.  
The highly correlated $d$- and $f$-electrons cannot be fully described by
effective single-particle methods, such as the local-density
approximation (LDA) to the density-functional theory (DFT). 
Here, the dynamical mean-field theory (DMFT) can be a powerful tool,
especially when the momentum dependence of the self-energy is
essentially negligible, regardless of the electron-electron interaction
strength.~\cite{springerlink:10.1007/BF01311397, PhysRevLett.62.324,
  RevModPhys.68.13}

The central problem of the DMFT is to solve an effective impurity
model.  
In real materials, such a model usually contains both inter- and
intra-orbital interactions, as well as the hybridization among
different orbitals. 
They account for the competitions between the magnetic, charge, and
orbital fluctuations. 
Thus, an efficient impurity solver, which can handle all the
interactions and hybridizations, is of obvious importance.  
Among the available impurity solvers,~\cite{PhysRevLett.72.1545,
  PhysRevB.45.6479, PhysRevLett.25.672, PhysRevB.36.2036,
  ASSP.46.169} the numerically exact quantum Monte Carlo (QMC)
methods were widely used. 
The recent development of the continuous-time quantum Monte Carlo
(CT-QMC) methods\cite{PhysRevB.72.035122, EPL.82.57003,
  PhysRevLett.97.076405, PhysRevB.74.155107}  
further supports the DMFT for the study of realistic materials in
the sense that lower temperature regions can be reached and more
orbitals can be investigated. 

For realistic material calculations based on the CT-QMC solvers,
correctly resolving the high-frequency behavior of  
the impurity self-energy $\Sigma_{imp}(i\omega_{n})$ is of crucial importance. 
On the one hand, due to the iterative nature of the DMFT equations,
$\Sigma_{imp}(i\omega_{m})$ determines the Weiss function at each
iteration and, in the end, the converged solution of the DMFT
procedure in some cases. 
On the other hand, $\Sigma_{imp}(i\omega_{n})$ strongly influences the
determination of the total particle number and the analytical
continuation for a full spectral function calculation, which has a
direct connection to experiments.  

In this paper, we show how to determine the impurity self-energy for a
rather general multiorbital model in an efficient and stable manner
within the hybridization-expansion continuous-time (CT-HYB) method.
The direct simulation in the Matsubara-frequency space and careful
treatment of the self-energy high-frequency tail make this method
especially suitable for studying the material systems with complex
band structures. 
 
This paper is organized as follow: Sec. \ref{Sec:Method} explains
how the ``dual transformation'' can be employed to simulate effectively 
the one-particle Green's function in the CT-HYB.  
Additionally, it is shown how the simulation of the two-particle
Green's function $\chi$ can be straightforwardly carried out as 
Wick's theorem still holds.  
The symmetry of $\chi$ is discussed in detail in this section.  
In Sec. \ref{Sec:Application}, we make use of the CT-HYB to study a
two-orbital Hubbard model with a general interaction term.  
For readers who are especially interested in our CT-HYB implementation
and the self-energy correction scheme, Sec. \ref{Sec:Method} is the
primary option.   
If the phase diagram of the two-orbital model is of primary interest, one
may skip Sec. \ref{Sec:Method} and go to Sec. \ref{Sec:Application},
which is self-contained. 
Conclusions and outlook can be found in Sec. \ref{Sec:Conclusion}.

\section{Method}\label{Sec:Method}

To explain our implementation of the CT-HYB in a concrete framework,
we take a two-orbital model as an example, that is,
\begin{eqnarray}\label{Eq:TwoOrbitalHubbardModel}
H_{loc} &=& H_{\Delta} + H_{V} + H_{int}
-\mu\sum_{a=1,2}\sum_{\sigma}n_{a\sigma} ,
\end{eqnarray}
where $H_{\Delta} = \Delta\sum_{\sigma}(n_{1\sigma}-n_{2\sigma})$ represents
the crystal field splitting and $H_{V} =
V\sum_{\sigma}(c_{1\sigma}^{\dagger}c_{2\sigma}+h.c.)$ is the
hybridization between two orbitals. 
For the interaction part, a general on-site form is considered,
\begin{eqnarray}
H_{int} &=& \sum_{a=1,2}Un_{a\uparrow}n_{a\downarrow} + \sum_{\sigma}
U^{\prime}n_{1,\sigma}n_{2,\bar{\sigma}}+\sum_{\sigma}U^{\prime\prime}n_{1\sigma}n_{2\sigma}\nonumber\\
&&-J(c_{1\downarrow}^{\dagger}c_{2\uparrow}^{\dagger}c_{2\downarrow}c_{1\uparrow}
+c_{2\uparrow}^{\dagger}c_{2\downarrow}^{\dagger}c_{1\uparrow}c_{1\downarrow} 
+ h.c.).
\end{eqnarray}
which contains the intraorbital and interorbital Coulomb
interactions, as well as the spin-flip and pair-hopping
processes.   

As an impurity solver for the DMFT, CT-HYB employs the same idea as
all the other CT-QMC impurity solvers; that is,
it expands the impurity effective action around a certain limit
and evaluates the expansion terms via stochastic sampling. 
Here, we only present the expressions relevant to this work. 
For a more detailed review of the CT-QMC methods, we suggest
Ref. \onlinecite{RevModPhys.83.349}. 

In the CT-HYB, the expansion of the ``impurity + bath" action
$S_{tot}=S_{loc}+S_{bath}+S_{hyb}$ around the atomic limit is carried
out by integrating out the bath degrees of freedom.  
$S_{loc}, S_{bath}$ are the actions for the local and the bath
Hamiltonian, respectively.  
$S_{hyb}$ is the hybridization between them,~\cite{RevModPhys.68.13}
which is expanded order by order.  
The contraction of the bath operator $b_{\sigma},
b_{\sigma}^{\dagger}$ follows Wick's theorem, as the bath is noninteracting.  
This results in a determinant $Det^{{\cal C}_{k}}$ with the
hybridization function $\Delta(\tau, \tau^{\prime})$ as matrix
elements. 
The full hybridization matrix usually can be decoupled into block
diagonal form with respect to certain conserved quantum numbers,  
for example, the total particle number $n$, the spin $\sigma_{z}$ and
cluster momenta $K$. 
The final expression of the partition function can then be written as
\begin{equation}\label{MultiBand}
    {\cal Z} = {\cal Z}_{b}{\cal Z}_{loc}\prod_{a}
    \sum_{k_{a}}\int_{0}^{\beta}
    \prod_{i=1}^{k_{a}}d\tau_{i}d\tau_{i}^{\prime}~\mbox{Tr}({\cal
      C}_{k_{a}}) Det^{{\cal C}_{k_{a}}} .
\end{equation}
Here, $k_{a}$ is the expansion order (also the dimension of the
determinant matrix) for the ``a" flavor, where flavor represents spin,
orbital, or cluster momenta.  
$\mbox{Tr}({\cal C}_{k})= \langle
T_{\tau}\prod_{a}c_{a}(\tau_{1}^{\prime})c_{a}^{\dagger}(\tau_{1})\cdots
c_{a}(\tau_{k}^{\prime})c_{a}^{\dagger}(\tau_{k})\rangle$ is the
cluster trace of a group of ``kinks",~\cite{PhysRevB.74.155107}
that is, cluster operators, in the interval $[0, \beta)$.
From now on, we always work with the diagonal form of the
hybridization function. 
The evaluation of $\mbox{Tr}({\cal C}_{k_{a}})$ can be carried out in
two ways. 
One can either express the $c_{\sigma}, c_{\sigma}^{\dagger}$
operators as matrices in the eigenbasis of $H_{loc}$ or employ the
Krylov implementation~\cite{PhysRevB.80.235117}.  
The former one benefits from the diagonal form of the time evolution
operators $e^{-H_{loc}\tau}$.   
The Krylov implementation, on the other hand, works in the
particle-number basis, for which $e^{-H_{loc}\tau}$ becomes a sparse
matrix.   
It uses the efficient Krylov-space method, which makes it possible to
simulate up to typically seven orbital problems at acceptable
numerical costs.  
In this work, the first implementation is used, in which we
diagonalize $H_{loc}$ with respect to the conserved quantum
numbers.~\cite{PhysRevB.75.155113}
The trace of the fermion operators is evaluated by first searching for
nonzero overlap between different eigenstates with respect to the
group of the cluster operators. The nonzero trace is, then, calculated
along the trajectory found. 

\subsection{One-particle Green's function}\label{Sec:1G}
The impurity Green's function is obtained by removing one row and
column from the determinantal matrix.  
$G_{a}(i\omega_{n})$ simply relates to $M=\Delta^{-1}$
by\cite{PhysRevB.74.155107, PhysRevB.75.155113} 
\begin{equation}\label{Eq:Ga}
G_{imp}(i\omega_{n}) = -\frac{1}{\beta}\sum_{i,j}M_{i,j}e^{i\omega_{n}(\tau_{i}-\tau_{j})}
\end{equation}
Alternatively, one can simulate the impurity Green's function from the
cluster trace at each Monte Carlo step;~\cite{PhysRevB.75.155113} that
is,  
\begin{eqnarray}
&&G_{imp}(i\omega_{n})=\frac{1}{\beta}\int^{\beta}_{0}e^{i\omega_{n}\tau}\langle
  c(\tau)c_{}^{\dagger}\rangle d\tau\nonumber\\ 
&=&\frac{1}{\beta{\cal Z}}\int^{\beta}_{0} d\tau
  e^{i\omega_{n}\tau}\sum_{\phi}\langle \phi\vert  
  e^{-\beta E_{\phi}} c(\tau)c_{}^{\dagger}\vert \phi\rangle 
\end{eqnarray}
Here, $\vert\phi\rangle$ is the eigenstate of the Anderson impurity
model, in terms of which the full partition function can be written as
${\cal Z}=\sum_{\phi}e^{-\beta E_{\phi}}$. 
For each specific configuration ${\cal C}_{k}$ sampled in the CT-HYB,
this expression has the following form: 
\begin{eqnarray}
G_{imp}^{{\cal C}_{k}} &=& \frac{{\cal Z}_{loc}{\cal
    Z}_{b}}{\beta{\cal Z}}\int^{\beta}_{0}d\tau e^{i\omega_{n}\tau}  
Det^{{\cal C}_{k}}\nonumber\\  
&&\hspace{0.5cm}\times\sum_{m}\langle m\vert e^{-\beta E_{m}} T_{1}^{l}
c(\tau) T_{l+1}^{k} c_{}^{\dagger}\vert m\rangle 
\end{eqnarray}
The explicit form of the determinant is given in
Eqn. (\ref{MultiBand}).  
$T_{1}^{l}$ and $T_{l+1}^{k}$ are the left and right lists of cluster
operators $c(\tau)$, respectively, with the constraint $\tau_{l+1} < \tau <
\tau_{l}$. 
The partition function corresponding to the configuration ${\cal
  C}_{k}$ is given as
\begin{eqnarray}
{\cal Z}_{{\cal C}_{k}}&=&{\cal Z}_{loc}{\cal Z}_{b}\sum_{m}\langle
m\vert e^{-\beta E_{m}} T_{1}^{l}\times T_{l+1}^{k}\vert m\rangle
Det^{{\cal C}_{k}}\nonumber\\  
&=& {\cal Z}_{loc}{\cal Z}_{b}\mbox{Tr}({\cal C}_{k}) Det^{{\cal C}_{k}}.
\end{eqnarray}
By combining the above two equations, we have
\begin{eqnarray}\label{Eq:FromClusterTrace}
G_{imp}^{{\cal C}_{k}}&=&\frac{{\cal Z}_{{\cal C}_k}}{\cal Z}
\frac{T}{\mbox{Tr}({\cal C}_{k})}\int^{\beta}_{0} d\tau
e^{i\omega_{n}\tau} \nonumber\\ 
&&\hspace{0.5cm}\times\sum_{m, l}\langle m\vert e^{-\beta E_{m}} T_{1}^{l}
c(\tau) T_{l+1}^{k} c_{}^{\dagger}\vert m\rangle
\nonumber\\ 
&=&\frac{{\cal Z}_{{\cal C}_k}}{\cal Z}  \frac{T}{\mbox{Tr}({\cal
    C}_{k})} \sum_{mn,pq,l}e^{-\beta E_{m}}T_{1,l}^{mn}c_{
  \omega}^{np} T_{l+1,k}^{pq}c_{}^{\dagger,qm}, 
\end{eqnarray}
with the notation $T_{1,l}^{m,n}\equiv \langle m \vert T_{1}^{l}\vert
n\rangle$ and  
\begin{equation}
c_{\omega}^{np}\equiv
\frac{e^{(i\omega_{n}+E_{n}-E_{p})\tau_{l}} -
  e^{(i\omega_{n}+E_{n}-E_{p})\tau_{l+1}} }
 {i\omega_{n} + E_{n}-E_{p}}\langle n \vert c\vert p\rangle.
\end{equation} 
The ratio ${\cal Z}_{{\cal C}_{k}}/{\cal Z}$ is the probability of
configuration ${\cal C}_{k}$ being sampled in the Monte Carlo
simulation. 

When $k_{a}$ is small, we measure $G_{imp}$ directly from the cluster
trace,~\cite{PhysRevB.75.155113}, that is, Eq. (\ref{Eq:FromClusterTrace}). 
Although this scheme is not very fast, it is more stable than
Eq.~(\ref{Eq:Ga}).  
When $k_{a}$ is large and Eq.~(\ref{Eq:Ga}) is used in the simulation, 
the high-frequency parts of $G_{imp}$ converge much slower and
contains more statistical errors than the low-frequency parts.  
As a result, the corresponding self-energy can be fluctuating at
large $\omega_{n}$. 
As already pointed out in the Introduction, the correct high-frequency
behavior of $\Sigma_{imp}(i\omega_{n})$ is crucial for the CT-HYB.  
Thus, special attention has to be paid to get rid of the noises in the
self-energy data. 

To the best of our knowledge, three schemes have been
proposed for dealing with this problem. (1) Noise filtering.  
One can either smooth the noises at $\tau\approx \beta/2$ by averaging
$G_{imp}(\tau)$ over a small range of $\tau$ (see
Refs. \onlinecite{PhysRevLett.97.076405, PhysRevB.74.155107})   
or apply the orthogonal polynomial filtering routine recently proposed
by Boehnke {\it et al}.~\cite{PhysRevB.84.075145} to achieve a smooth
$G_{imp}(\tau)$ for all $\tau \in [0, \beta)$. 
By carefully choosing the order of the orthogonal polynomials, the
impurity self-energy becomes smooth for all Matsubara frequencies.  
(2) Replacing the high-frequency tail of $\Sigma_{imp}(i\omega_{n})$
with some well-behaving function.  
This function can be either the self-energy, calculated from a
weak-coupling perturbation expansion,  
or the moment expansion of the Green's function.~\cite{Gull,
  PhysRevB.84.073104}
Such a replacement provides a smoothly behaving high-frequency tail of
the self-energy function.  
However, the corresponding expression usually becomes complicated in the
multi-orbital case and relies on the explicit form of the interaction
term. 
(3) Measuring $G_{imp}(\tau)$ from higher order
correlation functions.~\cite{2011arXiv1108.1936H} 
This method becomes advantageous for the density-density type
interaction, for which the ``segment
picture"~\cite{PhysRevLett.97.076405} can be used.     
For general type interactions, numerical cost has to be paid to calculate
additional correlators. 

Here, we propose a simple and stable scheme which does not
rely on any direct noise filtering of $G_{imp}(\tau)$  
and does not introduce any numerical cost to the CT-HYB simulations.  
This method does not depend on the explicit form of the interaction
term and remains efficient in the multiorbital calculations.  
The basic idea is to determine an approximate self-energy function by
performing the perturbation expansion around the atomic limit, using
the 'dual-transformation'.  
As we will see later on, such a method generates systematic improvements
to the atomic self-energy.  
The first-order expansion term already gives considerable corrections
and reproduces the correct high-frequency behavior of
$\Sigma_{imp}(i\omega_{n})$. 

The expansion around the atomic limit has been
studied before.~\cite{PhysRevB.72.045111}
In the strong-coupling region, this method yields results comparable
to the numerical exact QMC results.  
Here, we use an elegant and different way, that is, the ``dual
transformation".~\cite{PhysRevB.79.045133}
This transformation has been used in the construction of the
dual-fermion (DF) method,  
which gives an action well behaving in both the weak- and
the strong-coupling limits.  
Thus, our perturbation expansion actually also works in the
weak-coupling region. 

The impurity model has the following action:
\begin{equation}\label{Eq:c-action}
S[c^{*}, c] = S_{imp}[c^{*}, c] + \sum_{n}\sum_{a}c_{a}^{*}\Delta_{a}(i\omega_{n})c_{a}
\end{equation}
In the ``dual transformation", new variables $f^{*}, f$ are introduced
to rewrite the hybridization term in the following way: 
\begin{eqnarray}
&& e^{c_{a}^{*}\Delta_{a}(i\omega_{n})c_{a}}\det[\frac{\Delta_{a}}{\alpha^{2}}]^{-1} \nonumber\\
&=&\int
  e^{-\alpha(c_{a}^{*}f_{a}+f_{a}^{*}c_{a})-\frac{\alpha^{2}}{\Delta_{a}(i\omega_{n})}f_{a}^{*}f_{a}}{\cal
    D}[f^{*},f]. 
\end{eqnarray}
The complex number $\alpha$ can be arbitrary in the above
expression. In Ref.~\onlinecite{PhysRevB.79.045133},  
it is taken as the impurity Green's function. This makes the
correlator of the dual variables, that is,
$G^{d}=-\langle f_{a}f_{a}^{*}\rangle$, behaves like the one-particle
Green's function, which decreases as $1/i\omega_{n}$ for large
$\omega_{n}$. 
For simplicity, we take $\alpha$ as one. Although in this case, the
dual variables can not be interpreted as fermions, the impurity
Green's function remains the same.    

Integrating out the $c$ variable, the full action becomes a functional
which only depends on variables $f^{*}, f$, that is,
\begin{eqnarray}\label{Eq:f-action}
{\cal Z}&=&{\cal Z}_{f}{\cal Z}_{b}\int{\cal D}[f^{*},
  f]e^{-\sum_{a}f^{*}_{a}\Delta_{a}^{-1}f_{a}}\nonumber\\ 
&&\int{\cal D}[c^{*}, c]e^{-S_{imp}[c^{*},
    c]}\sum_{k_{a}}\frac{1}{k_{a}!}(c_{a}^{*}f_{a}+f_{a}^{*}c_{a})^{k_{a}}\nonumber\\ 
&=&{\cal Z}_{f}{\cal Z}_{b}\int{\cal D}[f^{*},
  f]e^{-\sum_{a}f^{*}_{a}G_{d}^{0, -1}f_{a} -  V^{(4)}_{d}f_{1}f_{2}^{*}f_{3}f_{4}^{*}}, 
\end{eqnarray}
where $G^{d}_{0}$ is given as $[G^{at}_{a} - \Delta_{a}^{-1}]^{-1}$.
The effective interaction of dual variables turns out to be the
reducible four-point correlations of the atomic system,  
that is, $V^{(4)}_{d}=\chi^{at}_{12;34}-\delta_{1,2}\beta
G^{at}_{12}G^{at}_{34} + \delta_{14}\beta G^{at}_{14}G^{at}_{23}$,
with $G^{at}$ being the atomic Green's function. 

Since the dual transformation is mathematically exact, the two
different actions which depend on only $c$
variables[i.e., Eq. (\ref{Eq:c-action})] and $f$variables
[i.e., Eq. (\ref{Eq:f-action})] are equivalent.   
Thus, we can obtain an exact relation between the correlators
$G_{a}$ and $G^{d}_{a}$ from differentiating the two actions with
respect to $\Delta_{a}$. This yields:
\begin{equation}\label{Eq:c-f}
G_{a}=-\Delta_{a}^{-1} - \Delta_{a}^{-1}G_{a}^{d}\Delta_{a}^{-1}, 
\end{equation} 
where $G^{d}_{a}$ is obtained from the Dyson equation, that is, $G^{d}_{a} =
[G_{0}^{d} - \Sigma^{d}_{a}]^{-1}$.  
$\Sigma^{d}_{a}$ is the self-energy function of the dual variables.  
The expression of $\chi_{12;34}^{imp}$ can be found in the
literature, (e.g., Refs. \onlinecite{springerlink:10.1134/S0021364010060123,
  PhysRevB.75.045118,0295-5075-85-2-27007}). 
If the interaction of the dual variables in Eq. (\ref{Eq:f-action}) is
neglected, the atomic self-energy will be recovered.  
This can be seen by inserting $G^{d}_{0}$ into Eq. (\ref{Eq:c-f}).
We have 
\begin{equation}
G_{a}(i\omega_{n}) = G^{at}_{a}/(1 - \Delta_{a}G^{at}_{a}). 
\end{equation}
Then, from the Dyson equation, we immediately see that
\begin{eqnarray}
\Sigma_{a}^{imp}&=& i\omega_{n}+\mu - \Delta_{a} - G_{a}^{-1}(i\omega_{n})\nonumber\\
&=&i\omega_{n}+\mu - 1/G^{at}_{a} = \Sigma_{a}^{at}
\end{eqnarray}
Thus, one can imagine the interaction term in Eq. (\ref{Eq:f-action})
will generate systematic corrections to the atomic self-energy. 

By including the interaction and further restricting the calculation
of $\Sigma^{d}_{a}$ to the first order, we have 
\begin{equation}
\Sigma^{d}_{a,\sigma}(i\omega_{n}) =
-\frac{1}{\beta}\sum_{b}\sum_{\omega_{n}^{\prime}}
V^{d,(4)}_{aa;bb}(i\omega_{n};i\omega_{n}^{\prime})G^{d}_{b}(i\omega_{n}^{\prime})  
\end{equation}
In this equation, only the element
$V^{d,(4)}_{12;34}\delta_{12}\delta_{34}$ is required.  
Additionally, this calculation can be further accelerated by employing
the look-up routine and the symmetry of $\chi_{12;34}^{at}$,  
which is shown in Sec. \ref{Sec:Gamma}.
By doing so, the perturbation expansion remains very efficienct in
multi-orbital calculations. 

As a benchmark, we first apply the dual expansion scheme by restudying
the Bethe lattice with different bandwidths, that is, $W_{2} = 2W_{1}$,  
where the orbital-selective Mott transition can
happen.~\cite{refId0, PhysRevLett.92.216402, PhysRevLett.95.206401,
 PhysRevB.72.205124, PhysRevLett.97.076405,PhysRevLett.99.236404,
 PhysRevLett.95.116402, PhysRevB.72.085112, PhysRevLett.91.226401,
 0953-8984-19-43-436206, PhysRevB.73.155106, PhysRevB.72.081103}   
We directly solved the DMFT equation with the high-frequency
supplemented self-energy function, instead of using Eq. (20) in
Ref. \onlinecite{PhysRevB.74.155107}. 
\begin{figure}[htbp]
\centering
\includegraphics[width=\linewidth]{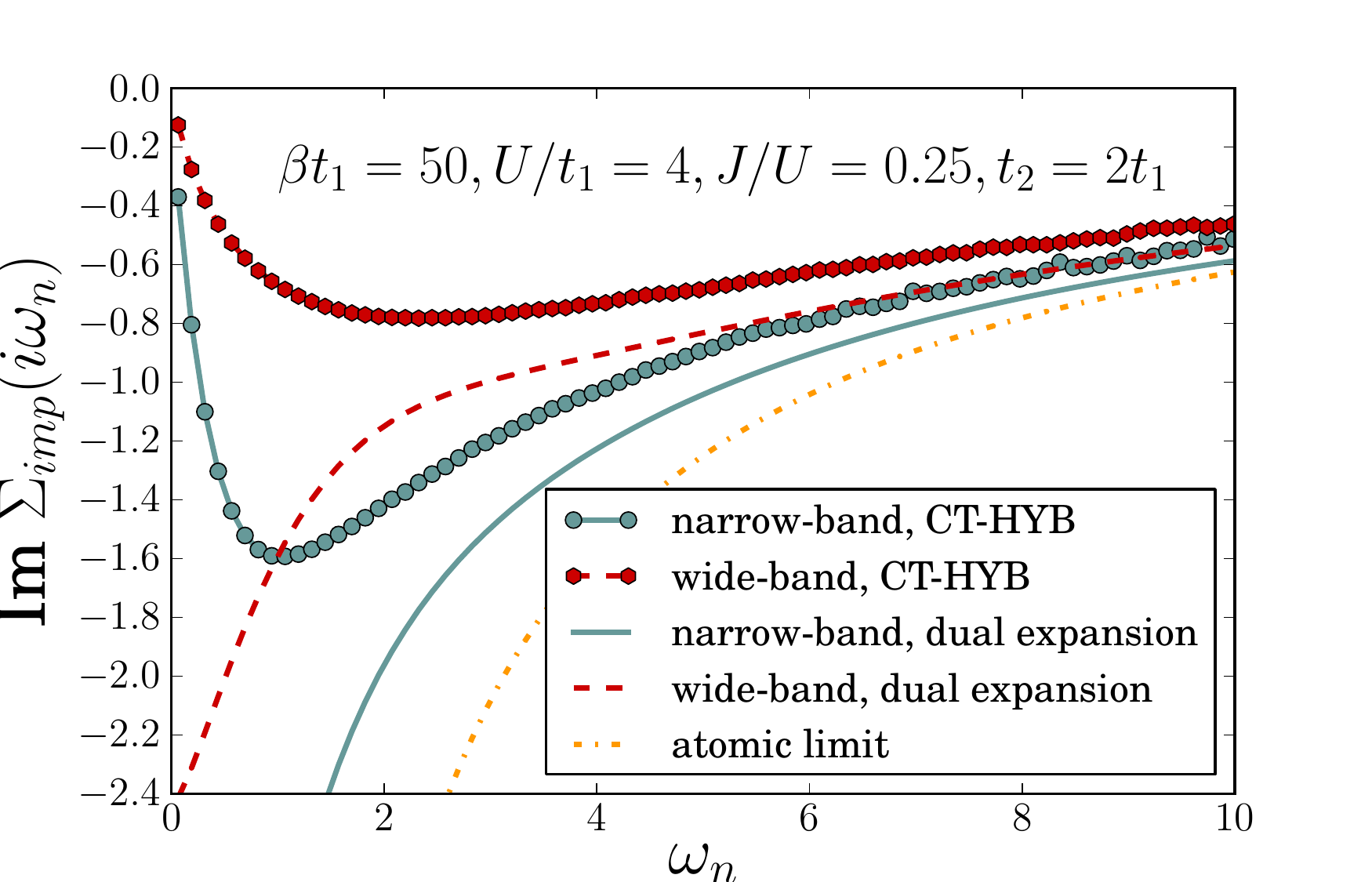}
\caption{Benchmark: imaginary part of the impurity self-energies for
  $\beta t_{1}=50, U/t_{1}=4, J/U=0.25$ in unit of $t_{1}$. Both the
  narrow and the wide bands are metallic.  
  The dual expansion gives two different asymptotic behaviors of the
  self-energy for two different bands, as expected.  
  However, the atomic self-energy does not have such a resolution.}
\label{Fig:Betheb50u4}
\end{figure}
Our self-energy data in Fig. \ref{Fig:Betheb50u4} is identical to those
in Fig. 12 of Ref. \onlinecite{PhysRevB.74.155107},  
meaning that the dual expansion method is reliable to produce the
high-frequency tail of the self-energy and can be used in the CT-HYB
for solving impurity problems. 
To see the performance of the dual expansion method for a finite
spatial-dimension problem, in Fig. \ref{Fig:DualExpansion}  
we show the comparison of the self-energy function calculated for a
two-orbital Hubbard model in two dimension [see the Hamiltonian in
Eq. (\ref{Eq:TwoOrbitalHubbardModel})].   
The improvement from the dual expansion is clearly seen from the
agreement between the CT-HYB and the dual expansion results. 
Increasing the hybridization strength, this agreement becomes even better.
Thus, a smaller number of Matsubara frequencies is required to
simulate in such a case.  
However, the atomic self-energy has a larger deviation from the CT-HYB
results for smaller $\omega_{n}$.  
Similar ideas were used to formulate effective impurity
solvers\cite{springerlink:10.1134/S0021364010060123,
  0295-5075-85-2-27007} for the DMFT.  
We use it here to get the correct high-frequency tail of the impurity
self-energy, while still keeping the low-frequency self-energy function
simulated from the QMC. 
This method only needs the hybridization function at each DMFT
iteration. 
The dual-expansion can be carried out independently of the
CT-HYB simulation. 
Thus, it does not introduce additional numerical cost to the CT-HYB,
which is another essential difference with respect to previous
works.~\cite{Gull, PhysRevB.75.155113, PhysRevB.84.073104,
  PhysRevB.84.075145, 2011arXiv1108.1936H} 
\begin{figure}[htpb]
\centering
\includegraphics[width=\linewidth]{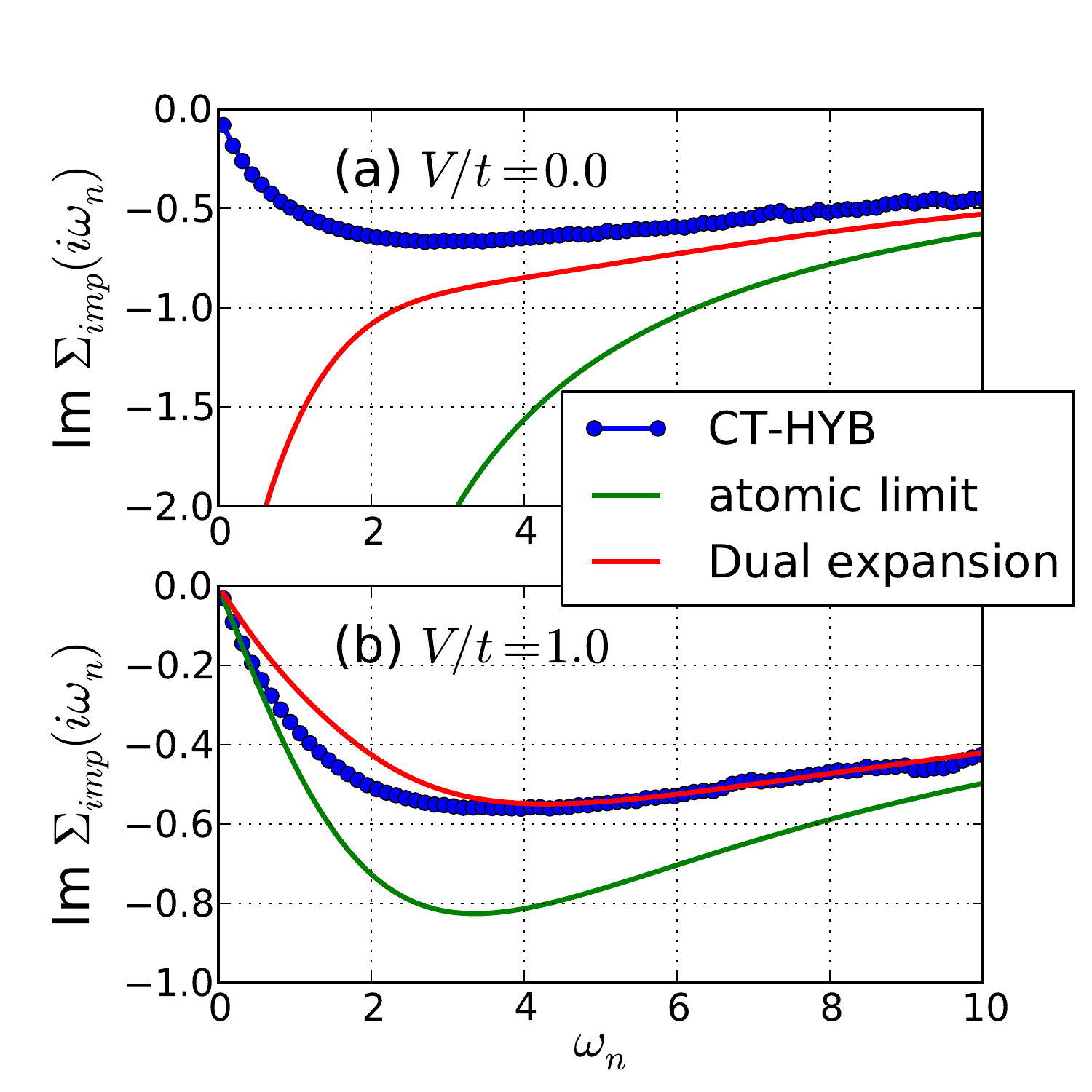} 
\caption{The comparison of the impurity self-energy calculated from
  the CT-HYB, atomic Hubbard model and the dual expansion method. The
  parameter sets for the two-orbital Hubbard model are $\beta t = 50,
  U/t = 4.0, J/U = 0.25, V/t = 1.0$. See the text for more details.} 
\label{Fig:DualExpansion}
\end{figure}

\subsection{Four-point correlation function
  \texorpdfstring{$\chi_{12;34}^{}$}{}}\label{Sec:Gamma} 
The dual expansion, discussed in the above section, requires the
knowledge of the atomic four-point correlation function
$\chi_{12;34}^{at}$.  
In the multiorbital case, such a calculation can be hard since the
large-dimensional matrix multiplication is time consuming.  
In this case, one can again use the block diagonal form of the
Hamiltonian matrix and employ the look-up routine as we did in the
trace calculation. 
Here, we want to further simplify the calculation by employing the
symmetry of $\chi_{12;34}^{at}$. 
Such a symmetry turns out to be also very useful in the simulation of the
impurity four-point correlation function $\chi^{imp}_{12;34}$. 
Thus, in this section we try to keep our discussion general. We start
from the simulation of the $\chi^{imp}_{12;34}$ in the CT-HYB  
and discuss the symmetries of it afterward. 
The same symmetry requirements are satisfied by $\chi^{at}_{12;34}$
as well. 

Although in the CT-HYB, Wick's theorem apparently is not supported by
the impurity action,  the four-point correlation function can be
simulated by removing two rows and two columns from the determinant
matrix, which results in an expression analogous to those for the
CT-INT and the CT-AUX. 
Effectively, one can still simulate the four-point correlation function
as if Wick's theorem holds. 
Here, we use the following notation to symbolically represent this
expression:
\begin{eqnarray}
  &&\chi^{}_{12;34} := \langle c^{}_{1}c^{*}_{2}c_{3}^{}c_{4}^{*}\rangle \nonumber\\
  &=& \overline{\langle c^{}_{1}c_{2}^{*}\rangle\langle
    c_{3}^{}c_{4}^{*}\rangle}- \overline{\langle
    c^{}_{1}c_{4}^{*}\rangle\langle c_{3}^{}c_{2}^{*}\rangle}
  \nonumber\\ 
  &=&\overline{g_{12}^{}(\omega^{}_{1},\omega^{}_{2})
    g_{34}^{}(\omega^{}_{3},\omega^{}_{4})}
  - \overline{g_{14}^{}(\omega^{}_{1},\omega^{}_{4})
    g_{32}^{}(\omega^{}_{3},\omega^{}_{2})}, 
\end{eqnarray}
where labels $12;34$ represent ``orbitals, sites, spins,'' etc.  
In the CT-HYB, the two-frequency dependent propagators
$g_{\alpha\beta}(\omega,\omega^{\prime})$ is given as 
\begin{eqnarray}
g_{\alpha\beta}(\omega_{1},\omega_{2}) =
-\frac{1}{\beta}\sum_{i,j}e^{i\omega_{1}\tau_{i}}M^{\alpha\beta}_{ij}e^{-i\omega_{2}\tau_{j}} .
\end{eqnarray}
It has the following symmetry in Matsubara frequency space:
\begin{equation}\label{Eq:symm-one}
g_{\alpha\beta}^{}(\omega_{1},\omega_{2}) = g_{\alpha\beta}^{*}(-\omega_{1},-\omega_{2}),
\end{equation}
which reduces the numerical effort by a factor of two. 
A similar symmetry is also satisfied by $\chi$:  
\begin{equation}\label{Eq:symm-two}
\chi_{12;34}=\chi^{\Omega}_{12;34}(\omega,\omega^{\prime}) =
\chi^{-\Omega,*}_{12;34}(-\omega,-\omega^{\prime}). 
\end{equation}
In what follows, we denote $\omega=\omega_{1}$,
$\omega+\Omega=\omega_{2}$, $\omega^{\prime}+\Omega = \omega_{3}$,
$\omega^{\prime}=\omega_{4}$. 
Equation (\ref{Eq:symm-two}) says, only for $\Omega > 0$, $\chi$ needs to
be simulated.  

Symmetry (\ref{Eq:symm-two}) relates the positive frequencies to the
corresponding negative frequencies of $\chi$. 
It is also possible to find symmetries which connect different
$\omega$, $\omega^{\prime}$ in the same $\Omega$ sector. 
This can be achieved via the fact that 
$\chi_{12;34}$ is antisymmetric under the exchange
$1\Leftrightarrow3$ and $2\Leftrightarrow4$: 
\begin{eqnarray}
\chi_{34;12}(\omega^{\prime}+\Omega,\omega+\Omega;-\Omega) =
\chi_{12;34}(\omega,\omega^{\prime};\Omega) \label{Eq:symm-three}.
\end{eqnarray}
Combining Eq. (\ref{Eq:symm-three}) with Eq. (\ref{Eq:symm-two}), we
have
\begin{equation}\label{Eq:symm-four}
\chi_{34;12}(-\omega^{\prime}-\Omega,-\omega-\Omega;\Omega) =
\chi_{12;34}^{*}(\omega,\omega^{\prime};\Omega). 
\end{equation}
Given the spin configurations of different $\chi$ channels, we find
$\chi_{12;34}^{\sigma\sigma;\sigma\sigma}$ satisfies both symmetries in  
Eqs. (\ref{Eq:symm-two}) and (\ref{Eq:symm-four}). 
However, $\chi_{12;34}^{\sigma\sigma;\bar{\sigma}\bar{\sigma}}$ only
satisfies the symmetry shown in Eq. (\ref{Eq:symm-two}) and the
following relation:
\begin{equation}\label{Eq:symm-five}
\chi_{12;34}^{\sigma\sigma;\bar{\sigma}\bar{\sigma}}=\chi_{34;12}^{\bar{\sigma}\bar{\sigma};\sigma\sigma}.
\end{equation}

One can implement the symmetries in Eqs. (\ref{Eq:symm-two}) and
(\ref{Eq:symm-four}) as follows.  
(1) $\chi^{\uparrow\uparrow;\uparrow\uparrow}$,
$\chi^{\downarrow\downarrow;\downarrow\downarrow}$, and
$\chi^{\uparrow\uparrow;\downarrow\downarrow}$ are
simulated only for $\Omega>0$. 
(2) For each specific $\Omega$ considered, Eq. (\ref{Eq:symm-four}) is
further applied to $\chi^{\uparrow\uparrow;\uparrow\uparrow}$ and
$\chi^{\downarrow\downarrow;\downarrow\downarrow}$.  
Only for parts of the frequency points in this $\Omega$-sector do
$\chi^{\uparrow\uparrow;\uparrow\uparrow}$ and
$\chi^{\downarrow\downarrow;\downarrow\downarrow}$ need to be simulated. 
(3) At the end of the calculation, $\Omega < 0$ components are
calculated through Eq. (\ref{Eq:symm-three}). 
(4) $\chi^{\downarrow\downarrow;\uparrow\uparrow}$ is calculated by
Eq. (\ref{Eq:symm-five}).   
In addition to the symmetries shown in Eqs. (\ref{Eq:symm-two}) and
(\ref{Eq:symm-four}), it is possible to find more symmetries to relate
different frequency sectors.  

Before finishing this section, we want to note that
the four-point correlation function is useful not only for the physical
response function and the dual-expansion scheme,  
but also relates closely with the extension of the DMFT.
In the DF method\cite{PhysRevB.79.045133} and the
dynamical vertex approximation (D$\Gamma$A),~\cite{PhysRevB.75.045118}
the nonlocal self-energy is constructed from the impurity
two-particle vertices. 

\section{Application}\label{Sec:Application}
As a typical application, we consider here a two-orbital Hubbard model
[see the Hamiltonian in  Eq.~(\ref{Eq:TwoOrbitalHubbardModel})],  
 with rotationally invariant interactions, i.e. $U^{\prime} = U - 2J,
 U^{\prime\prime} = U^{\prime} - J$.  
 To make a link with realistic material systems, this 
multiorbital Hubbard model can be viewed as an effective model for
the $e_{g}^{}$-orbital systems.
The rotational invariance of the interaction term is not
obligatory in the CT-HYB solver; 
here, we use it only as one possible situation. 
By making use of the DMFT, the two-orbital Hubbard model has been
studied by many groups.~\cite{PhysRevB.77.125117, PhysRevLett.99.126405, 
  PhysRevB.80.155116, PhysRevB.83.125110,  
JPSJ.76.094712, PhysRevB.74.155107, Koyama20093267,
PhysRevB.83.205112, PhysRevLett.107.256401} 
These calculations are either based on a semicircular density of
states, which corresponds to the Bethe lattice,  
or they employ an impurity solver with certain limitations in
temperature or interaction strength. 
Here, we solve the DMFT equation at finite dimension and temperatures. 
In these cases, the DMFT loop cannot be closed by a simple relation in
the imaginary-time space like on the Bethe lattice. 
Thus, our dual-expansion method discussed in Sec.~\ref{Sec:1G} turns
to be a decisive tool.
Our calculations are mainly performed on ordinary desktop computers. 

\begin{figure}[t]
\centering
\includegraphics[width=\linewidth]{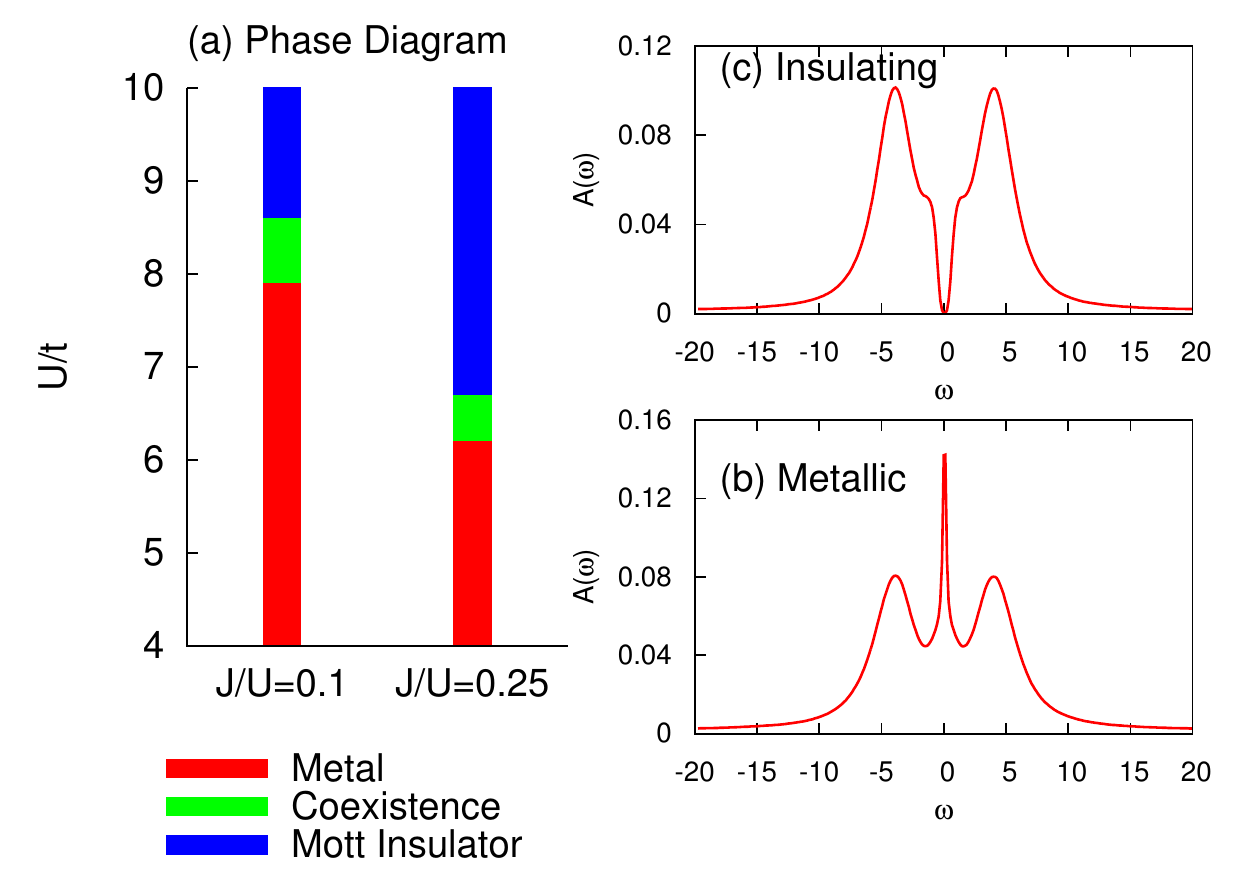}
\caption{The phase diagram of the two-orbital Hubbard model at half filling.
The MIT at $\beta t = 50$ for two values of $J/U$ are shown as histograms.
The two local density of states on the right-hand side correspond to
the two solutions for $U/t = 8.4, J/U = 0.1$.} 
\label{Fig:histo}
\end{figure}

Compared to the single-orbital case, two issues in a multiorbital
model are of obvious interests:\\
{\it (1) What is the effect of the orbital fluctuations? }
The general believe is, that it is competitive to the Coulomb interaction. 
As a result, the metallic state can be stabilized up to a large
interaction value\cite{PhysRevB.72.085112, PhysRevB.66.165107}. \\
{\it (2) How does the Hund's coupling modify the transition from the metal to
Mott insulator (MIT)? }
It is known that the two-orbital Hubbard model behaves quite
differently with and without $J$.~\cite{PhysRevB.77.125117,
  PhysRevLett.99.126405} \\
The phase diagrams of the two-orbital Hubbard model can be
found in Refs.~\onlinecite{PhysRevLett.99.126405} and
\onlinecite{PhysRevB.83.205112}.   
Here we study, in particular,  the coexistence region for different values
of $J$ in Fig. \ref{Fig:histo} (a), which indicates the MIT is of first
order. 
Compared to the phase diagrams for the Bethe
lattice,~\cite{PhysRevLett.99.126405, PhysRevB.83.205112}
the reduction of the spatial dimension does not change significantly
the critical Coulomb interaction value of $U_{c}$ when it is
normalized by the full bandwidth.   
However, $U_{c}$ becomes larger compared to the single-orbital model,
which confirms that the orbital fluctuation stabilize the metallic
phase.  
With the increase of the Hund's rule coupling $J$, we found the
coexistence region to become smaller.  
For the two values of $J/U$ in our calculations, the reduction is
about  0.2 eV.  
On the other hand, Bulla {\it et
al.}~\cite{springerlink:10.1140/epjb/e2005-00117-4} found, for 
$J/U > 0.25$,  the transition to be of second order. 
At $J/U = 0.25$,  our results show that the coexistence 
region still has a reasonably large width. Thus, we believe that
even for $J/U > 0.25$, the MIT remains first order.  
Whether, the coexistence region completely disappears with the further
increase of $J/U$ deserves more investigations.   

On the right-hand side of Fig. \ref{Fig:histo}, two different
solutions of the local density of state, that is, $A(\omega)$, are displayed
for $U/t = 8.4$.  
They correspond to the metallic, see Fig. \ref{Fig:histo} (b),  and
insulating states [see Fig.~\ref{Fig:histo} (c)] in the coexistence
region.   
$A(\omega)$ is obtained by using the stochastic analytical
continuation directly on the Matsubara data of
$G_{imp}(i\omega_{n})$.~\cite{2004cond.mat..3055B}

\begin{figure}[t]
\centering
\includegraphics[width=\linewidth]{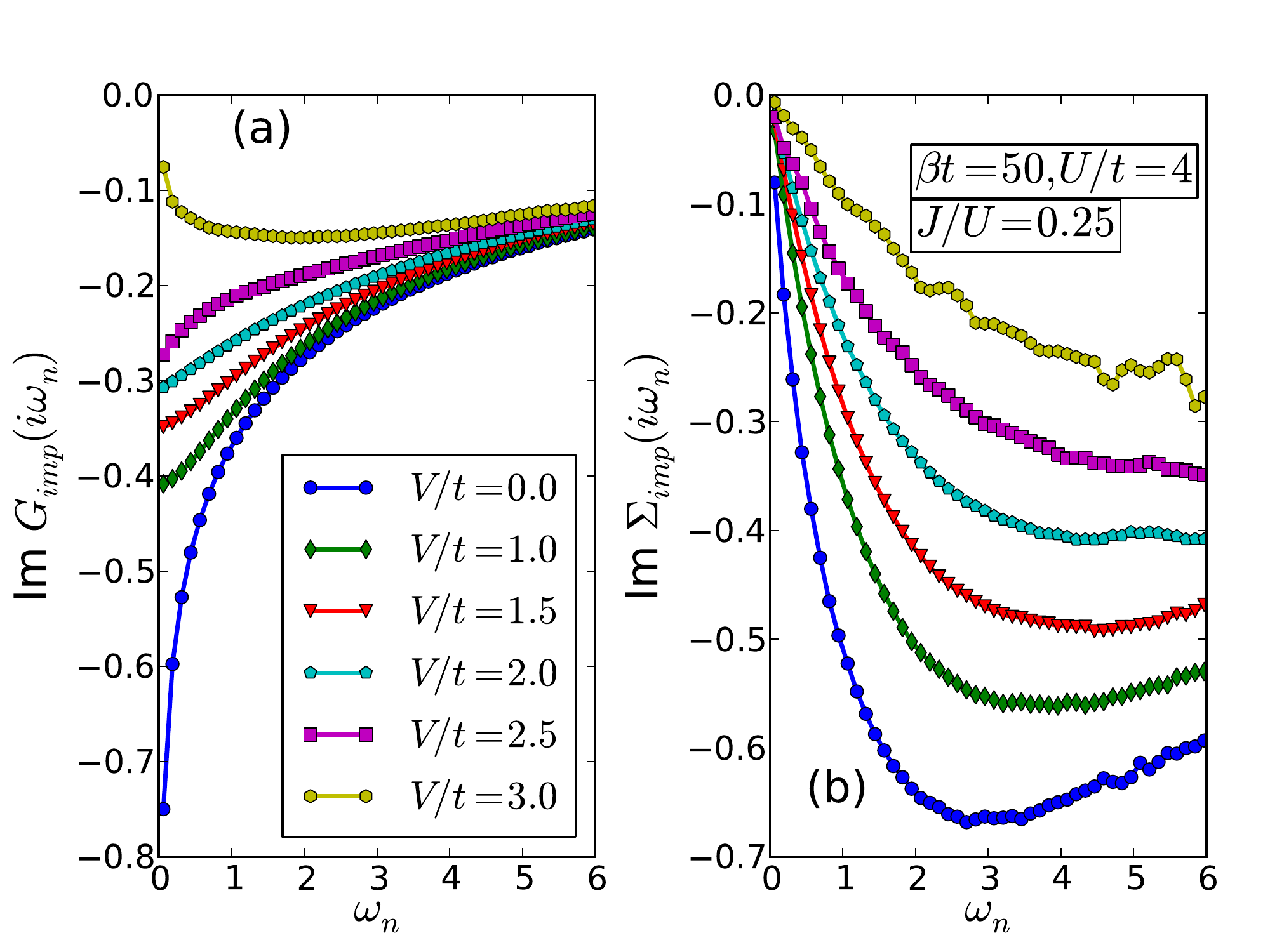}
\caption{Behavior of the impurity self-energy and Green's
  function around the metal-to-band-insulator transition as functions
  of the hybridization strength $V/t$. } 
\label{Fig:b50u4}
\end{figure}
In Fig. \ref{Fig:b50u4}, the typical behavior of the
metal-to-band-insulator transition is shown by calculating the
impurity Green's function and the corresponding self-energy as a
function of the hybridization.  
Increasing the hybridization $V/t$ tends to open a band gap. 
Furthermore, with the increase of $V/t$, 
the impurity Green's function at the lowest Matsubara frequency becomes
smaller and finally approaches zero [see Fig. \ref{Fig:b50u4} (a)].  
The metal-to-band-insulator transition happens somewhere between $V/t
= 2.5$ and $3.0$ for $U/t = 4$.  
This transition is not visible from the self-energy plot, where
$\Sigma_{imp}(i\omega_{n})$ behaves similarly for different values of
$V/t$. 
The slope, that is, $\partial \Sigma_{imp}(\omega)/\partial
\omega\vert_{\omega_{0}}$,  remains negative for all hybridization
strengths [see Fig. \ref{Fig:b50u4} (b)].   
In contrast, the slope of the local Green's function around
$\omega_{0}$ has different signs before and after the
metal-insulator (band) transition. 

\begin{figure}[b]
\centering
\includegraphics[width=\linewidth]{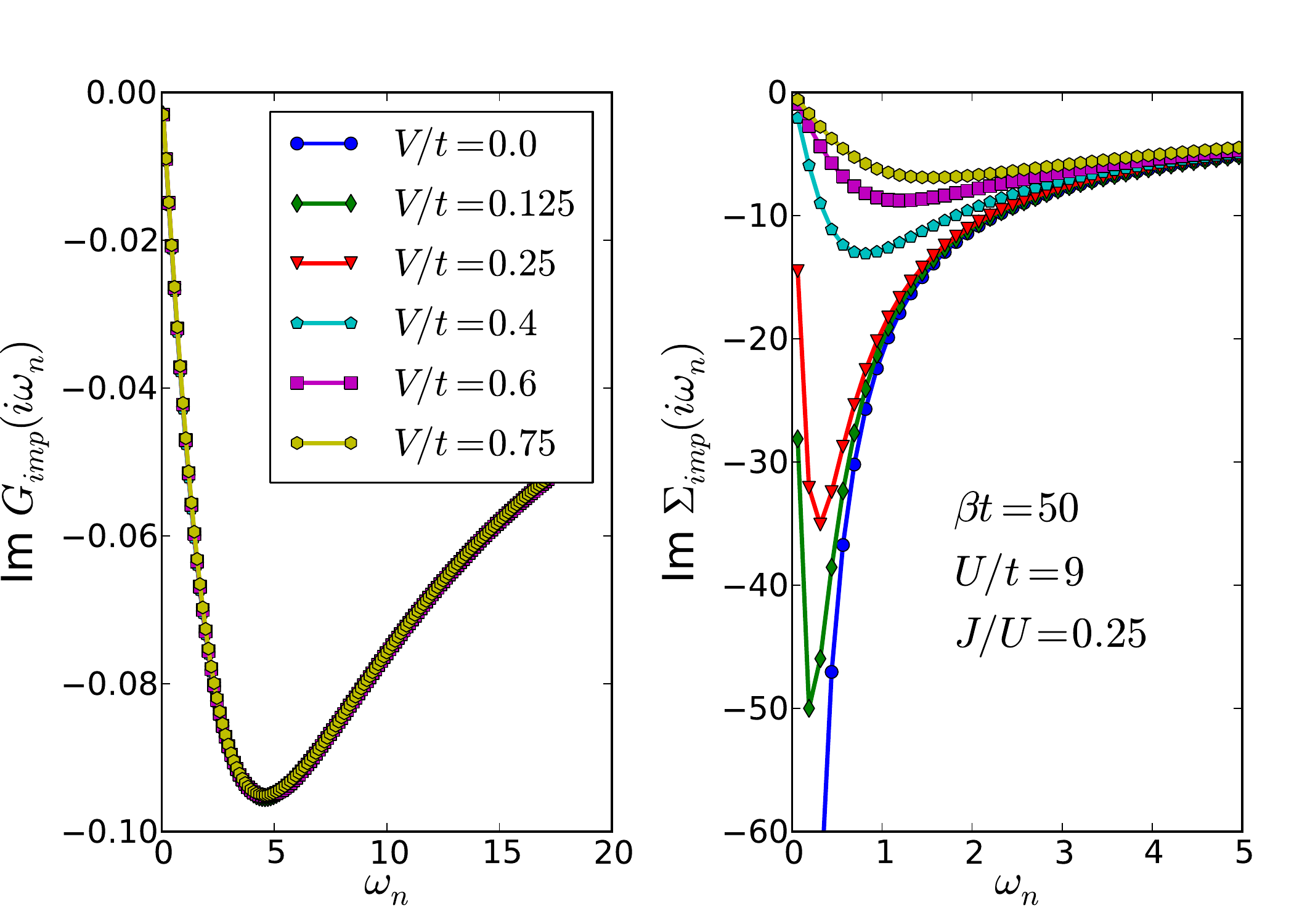}
\caption{Similar to Fig.~\ref{Fig:b50u4}, but with different
  interaction strength $U/t=9$, where a Mott-insulating state is found
  at $\Delta /t =0, 0.125, 0.25$. 
  The increase of the hybridization between two orbitals greatly changes
  the behavior of the self-energy, while it leaves the one-particle Green's
  function essentially unchanged.} 
\label{Fig:b50u9}
\end{figure}
Increasing further the value of $U/t$ strengthens both the intra- and
interorbital interactions.  
Finally, for values of $U/t$ of the order of the noninteracting
bandwidth, the metal to band-insulator transition is replaced by the
Mott-insulator-to-band-insulator crossover as a 
function of the hybridization strength $\Delta /t$.
This behavior is displayed in Fig. \ref{Fig:b50u9}. 
In contrast to the metal to band-insulator transition shown in
Fig. \ref{Fig:b50u4},  
in Fig. \ref{Fig:b50u9} (with a choice of $U/t=9$), the local Green's
function stays nearly unchanged under modifying the hybridization
strength $V/t$, that is, $G_{imp}$ shows an insulating behavior for all
values of $V/t$.   
However, for different values of $V/t$, the insulating nature is
indeed different.  
This can be seen from the variation of the self-energy function shown
in the right-hand side of Fig. \ref{Fig:b50u9}. 
Increasing $V/t$,  results in increasing of $\partial
\Sigma(\omega)/\partial \omega\vert_{\omega_{0}}$ for any finite
$V/t$,  
indicating the crossover from Mott-insulator to
band-insulator behavior.~\cite{PhysRevB.73.245118}

\begin{figure}[t]
\centering
\includegraphics[width=\linewidth]{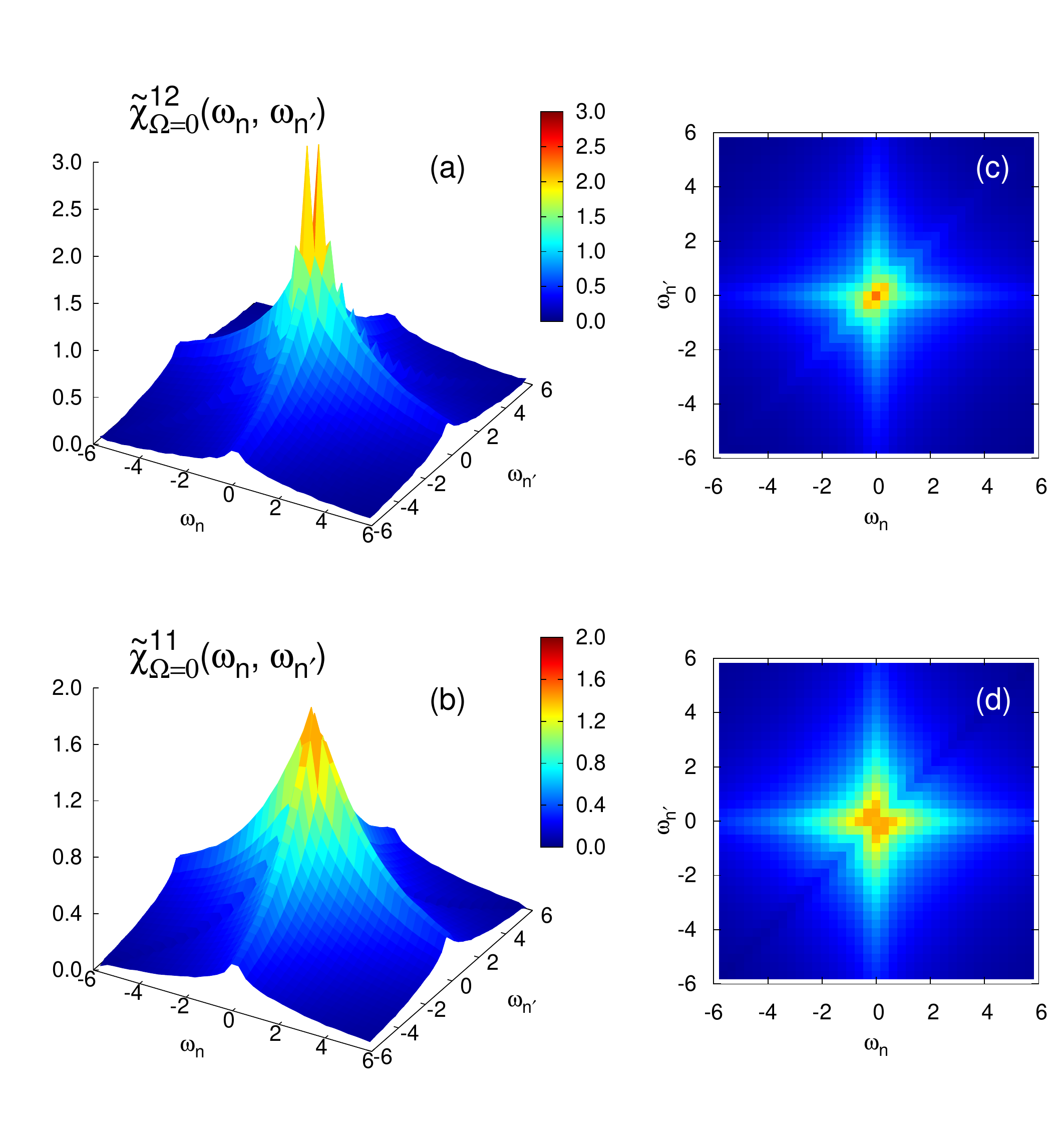}
\caption{The interorbital, that is, $\tilde{\chi}_{\Omega=0}^{12}$, and
  intraorbital, that is, $\tilde{\chi}_{\Omega=0}^{11}$, components of
  the reducible impurity two-particle susceptibility for $\beta t=20,
  U/t=6, J/U =0.25$.} 
\label{Fig:Gamma}
\end{figure}

By applying the symmetries presented in Sec. \ref{Sec:Gamma},  we show
the results for the interorbital and intraorbital reducible spin
susceptibilities in Fig. \ref{Fig:Gamma} for $\beta t = 20, U/t=6, J/U
= 0.25$, and $V/t=0$, with $a, b$ the orbital indices:   
\begin{equation}
\tilde{\chi}^{spin, ab}_{\Omega}(\omega_{n}, \omega^{\prime}_{n}) =
\frac{1}{2}(\tilde{\chi}_{ab,\Omega}^{\sigma\sigma,\sigma\sigma} -
\tilde{\chi}_{ab,\Omega}^{\sigma\sigma,\bar{\sigma}\bar{\sigma}}) 
\end{equation}

$\tilde{\chi}^{spin, ab}_{\Omega}(\omega_{n}, \omega^{\prime}_{n})$
are the impurity susceptibilities with the subtraction of the impurity
bubble susceptibilities.   
They are plotted as functions of the two fermionic frequencies
$\omega_{n}, \omega_{n}^{\prime}$ for fixed $\Omega=0$. 
While here only the $\Omega=0$ component is given, the
implementation discussed in Sec. \ref{Sec:Gamma} works for any value
of $\Omega$.  
Figures~\ref{Fig:Gamma}(a) and \ref{Fig:Gamma}(b) refer to the
three-dimensional (3D) plots of $\tilde{\chi}^{spin, 
  ab}_{\Omega}(\omega_{n}, \omega^{\prime}_{n})$;
the corresponding 2D top-view plots are shown in
Figs.~\ref{Fig:Gamma}(c) and \ref{Fig:Gamma}(d). 
Based on the CT-HYB, the four-point correlation functions were recently
also calculated for the effective one and four-orbital
systems\cite{PhysRevB.84.075145, PhysRevLett.107.137007} for different
problems.  
Another efficient and stable, but approximate, algorithm can be found
in Ref. \onlinecite{PhysRevB.83.085102}. 

From Fig. \ref{Fig:Gamma}, we see that the reducible two-particle
susceptibility $\tilde{\chi}_{\Omega=0}^{spin, ab}(\omega_{n},
\omega_{n^{\prime}})$ decays rather fast as a function of $\omega_{n}$
and $\omega_{n^{\prime}}$.  
The dominant contribution comes from the elements with $\omega_{n}=0$,
or $\omega_{n^{\prime}}=0$, or $\omega_{n}=\omega_{n^{\prime}}$.  
For our parameter set, the interorbital spin susceptibility shows a
sharper structure than the intraorbital one,  
which can be viewed as a precursor of the possible orbital
antiferromagnetic order.  

\section{Conclusion}\label{Sec:Conclusion}
In this paper, we showed how the high-frequency tail of the
self-energy can be calculated in a controlled manner from the dual
transformation in CT-HYB.  
This scheme provides an efficient recipe for finite-dimension DMFT
studies when taking the CT-HYB as an impurity solver.   
Our procedure is based on a  Matsubara frequency space simulation and 
produces more moments from the dual expansion.  
Thus, it generates an improved high-frequency self-energy behavior.   
Most importantly, it does not introduce any additional numerical cost
to the runtime simulation. 
We also simulated the four-point correlation function for different spin
configurations in the particle-hole channel. 
To this end, we implemented different symmetries to reduce the memory and
CPU requirements without losing accuracy.   

As a first application, we demonstrated the usefulness of our method
for a two-orbital model with a general on-site interaction.  
From this study, we deduced a substantial influence of the Hund's rule
coupling on the metal-insulator transition phase diagram, especially
on the coexistence region.  
In particular, we find that for any finite value of $J/t$, the MIT
stays first order.  

Our scheme is also of particular use for connecting the DF
method, which many be viewed as a nonlocal extension of the DMFT,
with {\it a priori} DFT techniques. 
A multiorbital DF calculation will be especially
interesting and rewarding for the DFT + DF study of material
systems.   
In such study, the CT-HYB effectively works on an impurity problem
with the DFT dispersions as input.  
Thus,  one has a good control on the "minus-sign" problem. 
The high-momentum resolution, provided by the DF algorithm, makes the
result ready to be compared with experiments, such as ARPES data.  

\acknowledgments
One of us (G. Li) acknowledges the valuable discussions with Philipp
Werner, Hartmut Monien, Xi Dai and Zhong Fang and is grateful for the
hospitality of Institute of Physics, Chinese Academy of Science.  
We thank Fakher Assaad for providing us the initial stochastic
analytical continuation code, from which the extension to Matsubara
frequency space was made.  
This work was supported by the DFG Grants No. Ha 1537/23-1 within
the Forschergruppe FOR~1162.

\bibliographystyle{apsrev4-1}
\bibliography{ref}
\end{document}